\magnification\magstep1

{\bf What is "system" : the arguments from the decoherence 
theory}

\bigskip

\centerline{Miroljub Dugi\' c}

\centerline{Faculty of Science, Department of Physics, P.O.B. 60}
\centerline{34 000 Kragujevac, Yugoslavia}

\centerline{E-mail: dugic@uis0.uis.kg.ac.yu}

\bigskip

{\bf Abstract:} Within the decoherence theory we 
investigate the physical background of the 
condition of the separability (diagonalizability in nonocorrelated
basis) of the interaction Hamiltnonian of the comopsite system,
"system plus environment". It proves that the condition of the
separability may serve as a criterion for defining "system",
but so that "system" cannot be defined unless it is simultaneously
defined with its "environment". When extended to a set of the
mutually interacting composite systems, this result implies that
the separability conditions of the local 
interactions are mutually tied. 
The task of defining "system" (and"environment")
via investigating the separability of the Hamiltonian is a sort
of the inverse task of the decoherence theory. A simple example of
doing the task is given.

\bigskip

{\bf PACS nuber: 03.65Bz}

\bigskip

{\bf 1. Introduction}

\bigskip

There is considerable interest in the {\it theory of decoherence},
and particularly in the "environment-induced superselection rules (EISR)"
theory [1, 2].This theory has been criticised by Machida and Namiki [3], 
particularly by posing the next question : "why environment would 
be so 'clever' as to recognize the 'pointer basis'?". This is really a 
substantial point in foundations of the EISR (decoherence) theory, for
appointing the specific role of the "environment" in this theory. Actually,
the role of the "environment" in the EISR theory is to meet some
requirements (e.g., to "recognize the 'pouinter basis' "), while itself being
an almost ill-defined quantum system (its degrees of freedom are usually
considered to be unobservable, while energetically the "environment"
is usually considered [4] to be equivalent with the "bath" of the harmonic
oscillators in the thermodynamical equilibrium). Thus, relative to the
(open quantum) "system (S)", the "environment (E)" is of the "secondary"
importance in the decoherence theory (its degrees of freedom always being
"traced out" in the corresponding calculations).

Recently [5, 6] it was pointed out existence of the necessary 
conditions for the occurrence of the "environment-induced
superselection rules" (decoherence). Rigorously speaking, 
these are the effective necessary conditions but (cf. Appendix I
below) one may forget about this "effectiveness".
The conditions are: (i) Separability of the interaction Hamiltonian
of the composite system "system plus environment (S+E)", $\hat H_{int}$,
and (ii) "Nondemolition" character of $\hat H_{int}$: $[\hat H_{int}(t),
\hat H_{int}(t')] = 0$. Let us emphasize (for more details see Section 2): 
the separability means that
there exists a (orthonormalized) basis in $H^{(S)}$ which diagonalizes
$\hat H_{int}$, {\it and} that there exists a basis in 
$H^{(E)}$ which also
diagonalizes $\hat H_{int}$; $H^{(S)}$ and $H^{(E)}$ represent the
Hilbert state spaces of the system (S), and of the environment (E), 
respectively. [If $\hat H_{int}$ proves nondiagonalizable in 
$H^{(S)}$ and/or in $H^{(E)}$, we say that such interaction Hamiltonian is 
of the nonseparable kind.] If any of the two conditions, (i) and/or (ii),
does not prove valid, one obtains nonoccurrence of decoherence, i.e.
{\it nonexistence of the "pointer basis"} of the system S. Therefore,
one obtains the answer to the question of Machida and Namiki : environment 
{\it needs not} to be so "clever" as to recognize the "pointer basis".

Still, this is a mathematical result which, by itself, hardly can
be considered physically very transparent. And this especially in 
the context 
of the Zurek's phrase [2] "no system no problem". Actually, this phrase
refers to the "requirement for classicality" [2], which considers 
nonoccurrence of decoherence (nonexistence of the "pointer basis")
as physically not very interesting issue.

Bearing this phrase in mind, in this paper we prepare an analysis of the 
condition of the separability. The analysis particularly referes to
the next question: whether the cannonical transformations in the composite 
system (S+E) can help in transforming  a nonseparable (separable)
interaction Hamiltonian into a separable (nonseparable) form (thus 
eventually
overcoming the predicted nonoccurrence of decoherence)? This way one 
obtains
some interesting results, while making connection to the problem [2]
"what is 'system'?"; by "system" we mean a given set of the "degrees of 
freedom". Actually, the analysis distinguishes the condition of the 
separability as a {\it criterion for defining "system"}, but so that
{\it "system" is not defined unless it is simultaneously defined by its
"environment"}. This is a new role of the "environment" in the 
EISR theory which has been anticipated by Machida and Namiki [3].

When extended to analysing a (macroscopic) system $S$, which consists in
many composite systems ($S = \cup_i (S_i+E_i)$), 
this result establishes that
the conditions of the separability of the
local interactions
in the system $S$ are {\it mutually tied}. This gives 
an interesting and consistent picture in the EISR theory which represents
a more rigorous formulation of the "requirement for classicality",
which is otherwise an intuitive and only plausible statement.

The plan of this paper is as follows. In Section 2 we give the different
definitions of the separability, which, as a necessary condition for 
decoherence, is confronted with the "requirement for classicality". In
Section 3 this situation is elaborated, thus leading to all the afore
mentioned results. Section 4 is discussion. Section 5 is conclusion.

\bigskip

{\bf 2. Nonseparability : nondivisability of "system" and "environment"}

\bigskip

In Ref. [6] it was proved that each (time independent) interaction 
Hamiltonian, $\hat H_{int}$, can (nonuniquely) 
be (re)written in the "linear" form:
$$\hat H_{int} = \sum_k \hat C_{Sk} \otimes \hat D_{Ek},
\eqno (1)$$

\noindent
but so that both sets of the observables, $\{\hat C_{Sk}\}$ of the
"system (S)", and $\{\hat D_{Ek}\}$ of the "environment (E)", consist
in {\it linearly independent observables}; i.e., $\sum_k \alpha_k
\hat C_{Sk}$ $= 0$ $\Rightarrow \alpha_k = 0, \forall{k}$, {\it and}
$\sum_k \beta_k
\hat D_{Ek} = 0 \Rightarrow \beta_k = 0, \forall{k}$. 

Along with the proof of existence of the form Eq.(1), it was developed a
method [6] for obtaining a particular form of $\hat H_{int}$ of
the type Eq. (1). This is a basis of the, so-called, "operational
definition" of the separability.

The next four {\it definitions of the separability} are mutually
equivalent :

(i) $\hat H_{int}$ is of the separable kind (i.e., it represents a 
separable interaction) if there exists a basis $\{\vert \phi_{Si}\rangle\}$
in the Hilbert state space of the "system", $H^{(S)}$, which diagonalizes
$\hat H_{int}$, {\it and} if there exists a basis $\{\vert \chi_{Ej}
\rangle\}$ in the Hilbert state space of the "environment", $H^{(E)}$,
which also diagonalizes $\hat H_{int}$.

(ii) $\hat H_{int}$ is of the separable kind if there exists a {\it 
noncorrelated} basis in the Hilbert state space of the composite system (S+E),
$\{\vert \phi_{Si}\rangle \otimes \vert \chi_{Ej} \rangle\}$, which
diagonalizes $\hat H_{int}$.

(iii) $\hat H_{int}$ is of the separable kind if its spectral form is of 
the next type :
$$\hat H_{int} = \sum_{p,q} \gamma_{pq} \hat P_{Sp}
\otimes \hat \Pi_{Eq}, \eqno (2)$$

\noindent
where $\gamma_{pq}$ represent the eigenvalues of $\hat H_{int}$, while
$\hat P_{Sp}$ and $\hat \Pi_{Eq}$ being the projecors onto the
subspaces of $H^{(S)}$ and of $H^{(E)}$, respectively.

(iv) ({\it the "operational definition"}) 
$\hat H_{int}$ is of the separable kind if and only if, for a
particular form of $\hat H_{int}$ of the type Eq. (1), one may state :
$$[\hat C_{Sk}, \hat C_{Sk'}] = 0, \forall{k,k'},
\eqno (3a)$$

\noindent
{\it and}
$$[\hat D_{Ek}, \hat D_{Ek'}] = 0, \forall{k,k'},
\eqno (3b)$$

If $\hat H_{int}$ is not of the separable kind, we say that it is of the
{\it nonseparable kind}.

Mutual equivalence of the first three definitions is rather obvious,
while their equivalence with the "operational definition" is
proved in Ref. [6]; the "operational definition"
will be of special interest below.

It is important to note that conclusion concerning the 
(non)separability
of a particular $\hat H_{int}$ uniquely and directly follows from
a particular form of $\hat H_{int}$ of the type Eq. (1), being
completely independent on the definitions of the observables
$\hat C_{Sk}$ and $\hat D_{Ek}$. Being a characteristic
of $\hat H_{int}$, the (non)separability puts speciffic limitations
on the possible forms of $\hat H_{int}$. For instance, for the separable
interaction, if one would obtain a "linear" form in which appear mutually 
incompatible observables of the "system" and/or of the "environment, it 
follows that the set(s) of the observables bears linear
dependence. Further, if $\hat H_{int}$ is of the nonseparable kind,
then {\it each form} of $\hat H_{int}$ bears incompatibility in the
set of the observables of the "system" and/or of the "environment".

As long as one is concerned with the time independent interactions,
the occurrence of decoherence relies only upon the condition of the
separability of $\hat H_{int}$. [Note : in general, for a time
dependent interaction, the expression Eq. (1) refers to a particular
instant, $t$.] Further, we shall be concerned only with the time independent
interactions. Finally, it is worth stressing that the separability
cannot be considered as a suficient condition for the occurrence of
decoherence. This is somewhat a more subtle issue, which here will
not be elaborated.

It is probably obvious (cf. Ref. [5]) that the nonseparability of 
$\hat H_{int}$ implies mutual {\it indistinguishability} 
({\it "indivisability}" [2]) of the "system (S)" and its 
"environment (E)".
That is, the nonseparable interactions in the composite system
S+E do not allow for putting a definite "border line" between
S and E. In the light of the Zurek's phrase [2] "no system no
problem", one might pose the question of physical relevance and 
usefulness of the notion of existence of the necessary
conditions for the occurrence of decoherence.

Actually, the requirement of divisability of S and E is the
"requirement for classicality" [2], without which "the problems with
the correspondence between quantum physics and the familiar everyday
classical reality cannot be even posed" [2]. In other words,
the requirement for classicality appears as a sort of a necessary
condition in the decoherence and in the quantum measurement theory.

Since the nonseparability {\it cannot meet this requirement}, one
may wonder if the nonseparability represents just a pathology
of the EISR theory, without any nontrivial physical meaning.
However, as it will be shown below, even in the context of the 
requirement for classicality (further : RC), the formal existence
of the nonseparable interactions provides us with some
interesting physical notions.

\bigskip

{\bf 3. Separability as a criterion for defining "system"}

\bigskip

Let us put RC in a more tractable form. For this purpose we are 
concerned 
with a system $S$, which is a set of (many-particle) quantum
systems, $S_i$ ($S = \cup_i S_i$). Each subsystem $S_i$ is an open
system, interacting with its environment, $E_i$. The Hamiltoniasn of 
the system $S$ is given by :
$$\hat H = \hat H_{\circ} + \sum_i \hat H^{(i)}_{int}
+ \sum_{i \neq j} \hat H^{(S)}_{ij} + 
\sum_{i \neq j} \hat H^{(E)}_{ij}, \eqno (4)$$

\noindent
where $\hat H_{\circ}$ represents the Hamiltonian of noninteracting
systems, $\hat H^{(i)}_{int}$ denotes interaction in the pair
$S_i+E_i$, while by the superscript "$S$" denoting the interactions
between the "systems" (for instance $S_i$ and $S_j$), and by the
superscript "$E$" denoting the interactions of the "environments" 
(for instance, of $E_i$ and $E_j$). {\it By definition}, each
interaction $\hat H^{(i)}_{int}$ is of the separable kind.

For this composite system, the RC can be formulated as follows :
{\it an interaction} $\hat H^{(i)}_{int}$, {\it which is of the
separable kind, cannot be changed into a nonseparable interaction,
either spontaneously, or by an action from outside}.

What is going to be shown is that RC can be {\it proved}, while 
providing us with some interesting notions within the 
decoherence theory.

\bigskip

{\bf 3.1 What is "system" ?}

\medskip

Let us go back to the expression Eq. (1). What is implicit in  
this expression 
is that each observable of the "system" and of the "environment" 
represents an analytical function of the corresponding degrees of
freedom; that is, $\hat C_{Sk} = C_k(\hat x_{Si}, \hat p_{Sj})$,
and $\hat D_{Sk} = D_k(\hat X_{E\alpha}, \hat P_{E\beta})$, where
the degrees of freedom satisfy $[\hat x_{Si}, \hat p_{Sj}] =
\imath \hbar \delta_{ij}$, and $[\hat X_{E\alpha}, \hat P_{E\beta}] =
\imath \hbar \delta_{\alpha \beta}$.

Now one may wonder if the cannonical (and particularly the linear)
transformations of the observables, $\hat x_{Si}, \hat p_{Sj},
\hat X_{E\alpha}, \hat P_{E\beta}$, can help in transforming $\hat 
H_{int}$ which is in a particular form
of the separable kind, into a form which is
of the nonseparable kind, and {\it vice versa}. Without any loss
of generality we shall further be concerned with the 
transformations of the nonseparable interactions.

At first glance, one obtains a straight answer to the above
question : since separability is a definite characteristic of
$\hat H_{int}$, one would expect that
it does not depend on either particular form,
or upon the choice of the degrees of freedom. However, this
answer is only partially correct.

Actually, validity of the above answer substantially depends
upon a sort of the cannonical transformations. Particularly, 
one may prepare the two different sorts of the cannonical
transformations : (a) the transformations which "mix" the
degrees of freedom of $S$, independently on the transformations
which "mix" only the degrees of freedom of $E$, and (b) the 
transformations "mixing" the degrees of freedom of both,
$S$ and $E$. So, the above answer refers
only to the transformations of the sort (a), which can be
proved as follows.

First, the cannonical transformations of this sort lead to the 
new degrees of freedom of the system $S$, $\{\hat \xi_{Sk},
\hat \pi_{Sl}\}$, independently on the degrees of freedom of 
the system $E$; let us by $\{\hat Q_{E\gamma}, \Pi_{E\delta}\}$
denote the "new" degrees of freedom of the system $E$. Then
the transformations of the sort (a) are presented by:
$$\hat \xi_{Sk} = \xi_k(\hat x_{Si}, \hat p_{Sj}),
\eqno (5a)$$
$$\hat \pi_{Sl} = \pi_l(\hat x_{Si}, \hat p_{Sj}),
\eqno (5b)$$
$$\hat Q_{E\gamma} = Q_{\gamma}(\hat X_{E\alpha},
\hat P_{E\beta}), \eqno (5c)$$
$$\hat \Pi_{E\delta} = \Pi_{\delta}(\hat X_{E\alpha},
\hat P_{E\beta}). \eqno (5d)$$

These transformations imply the transformations of the observables
appearing in Eq. (1) : e.g., $\hat C_{Sk} = C_k(\hat x_{Si},
\hat p_{Sj}) \rightarrow \hat C'_{Sk} = C'_k(\hat \xi_{Sk},
\hat \pi_{Sl})$, so giving rise to a new form of $\hat H_{int}$:
$$\hat H_{int} = \sum_m \hat C'_{Sm} \otimes \hat D'_{Em}.
\eqno (6)$$

According to the "operational definition" of the separability
one obtains: the nonseparability of $\hat H_{int}$ implies
existence of at least two observables of the system, $\hat C_{Sk}$
and $\hat C_{Sk'}$, for which $[\hat C_{Sk}, \hat C_{Sk'}] \neq 0$,
and/or analogously for the observables of the system $E$. Now, if 
the form Eq. (6) should be of the separable kind, the same
definition of the separability implies :
$$[\hat C'_{Sm}, \hat C'_{Sm'}] = 0, \forall{m,m'},
\eqno (7)$$

\noindent
{\it and} analogously for the observables of $E$.

However, the cannonical transformations of the sort (a) cannot
provide the loss of incompatibility which justifies the above 
statement.

On the other side, however, as regards the  transformations
of the sort (b), there is no a such obstacle in the same
concern. Actually, one can think of the transformations
of this sort, given by:
$$\hat \xi_{S'k} = \xi_k(\hat x_{Si}, \hat p_{Sj},
\hat X_{E\alpha}, \hat P_{E\beta}), \eqno (8a)$$
$$\hat \pi_{S'l} = \pi_l(\hat x_{Si}, \hat p_{Sj},
\hat X_{E\alpha}, \hat P_{E\beta}), \eqno (8b)$$
$$\hat Q_{E'\gamma} = Q_{\gamma}(\hat x_{Si}, \hat p_{Sj},
\hat X_{E\alpha}, \hat P_{E\beta}), \eqno (8c)$$
$$\hat \Pi_{E'\delta} = \Pi_{\delta}(\hat x_{Si}, \hat p_{Sj},
\hat X_{E\alpha}, \hat P_{E\beta}), \eqno (8d)$$

\noindent
so as to the nonseparable interaction can be transformed
into a separable form (and {\it vice versa}). {\it But this is a
substantial step, which is distinguished by the new subscripts},
$S'$ and $E'$. 

If possible at all (see Section 4 and Appendix II), 
these transformations
should lead to a new, separable form of $\hat H_{int}$,
but with respect to the {\it new sets of the degrees of 
freedom} : $\hat \xi_{S'k}, \hat \pi_{S'l}, \hat Q_{E'\gamma},
\hat \Pi_{E'\delta}$. Therefore, to be meaningfull, the transitions 
(8a-d) should define the new "system", $S'$, and its
(new) "environment", $E'$.

Thus one comes to the next notion : if possible at all, and if
meaningfull, the transformations (8a-d) lead to the {\it redefining
of the composite system}, $S+E$ (relative to whose degrees of freedom 
the given $\hat H_{int}$ appears to be of the nonseparable kind). Hence,
{\it instead of the "old" system} $S+E$, {\it one obtains a
new composite system}, $S'+E'$, whose degrees of freedom
define a separable form of the given interaction Hamiltonian,
$\hat H_{int}$. 

Once the new composite system is defined by the "new" degrees of
freedom, $(\hat \xi_{S'k}, \hat \pi_{S'l}; \hat Q_{E'\gamma},
\hat \Pi_{E'\delta})$, there remains tha task of the precise putting
the "border line (i.e., dividing this set onto the two subsets),
which gives {\it precise definition} of the "system" $S'$, and its 
"environment", $E'$. Although this needs not to be unique, it is 
important to stress that, as a matter of principle, it always can be
done. Actually, since the limit $N \to \infty$ is legitimate
($N$ being the number of "particles" in the "old" "environment"
$E$), the analogous limit is automatically fulfilled for the new
composite system (for no constraints of the degrees of freedom
have been involved). The task of putting the "border line"
is comparatively trivial in our considerations (this is just the
exchange of the "particles" in the "new" composite system), so 
further we shall assume this task completed. Thus one reaches the
point at which the application of the standard scheme of the EISR
theory is straightforward, which allows for defining the "pointer
basis" of the "system", $S'$.

Everything told in this subsection can be formally summarized as
follows: For a given set of the "degrees of freedom", 
$(\hat x_{Si}, \hat p_{Sj}; \hat X_{E\alpha}, \hat P_{E\beta})$,
an interaction Hamiltonian, $\hat H_{int}$, given in a particular form of the
type Eq. (1), is of the nonseparable kind. However, with respect to the
"new" set of the degrees of freedom, 
$(\hat \xi_{S'k}, \hat \pi_{S'l}; \hat Q_{E'\gamma}, \hat \Pi_{E'\delta})$, 
the same interaction
Hamiltonian obtains a separable form (of the type of Eq. (1)):
$$\hat H_{int} = \sum_p \hat E_{S'p} \otimes \hat F_{E'p},
\eqno (9)$$

\noindent
where $\hat E_{S'p} = E_p(\hat \xi_{S'k}, \hat \pi_{S'l})$, 
and $\hat F_{E'p} = F_p(\hat Q_{E'\gamma}, \hat \Pi_{E'\delta})$.

And this brings us to the next task : for an {\it a priori} given set
of the degrees of freedom, the nonseparability of a given interaction
Hamiltonian might be overcome by applying the cannonical transformations
of the sort (b), thus obtaining a definition of the "system"
(above : $S'$), {\it but only simultaneously} with obtaining
a definition of the corresponding "environment" (above : $E'$).
[Note: then the "old systems", $S$ and $E$, remain mutually 
indivisable.]

\bigskip

{\bf 3.2 The proof of RC}

\medskip

As regards the time independent interactions, the proof of RC 
relies upon the considerations of the actions from "outside"
the composite system $\cup_i (S_i+E_i)$.

The physical situation here to be analysed is the next one :
one wonders if by an action from outside, a local separable
interaction can be transformed into a nonseparable form. As
above, we shall be concerned with the inverse transformations,
bearing in mind the conclusion of subsection 3.1 : that such
transformations imply the redefining of an {\it a priori} given
composite system.

Let us refer to a particular composite system, $S_i+E_i$. Each 
action from outside assumes an interaction with an outer
quantum system, $A$. Since the task is to transform the
nonseparable, local interaction $\hat H_{S_iE_i}$ 
($\equiv \hat H^{(i)}_{int}$), into
a separable form, the system $A$ must interact with the
composite system $S_i+E_i$ as a whole; let us denote this
by $(S_i+E_i)+A$. Finally, being a macroscopic system, the
system $A$ is an open quantum system, thus leading to the
next physical situation: $(S_i+E_i)+A+E_A$, where $E_A$
denotes the environment of the system $A$.

It is important to note that, if the interaction of $A$ and the 
"whole", $S_i+E_i$, is of the separable kind, {\it everything
remains intact}. Actually, as it is implicit in the subsection
3.1, the separable interaction keeps the degrees of freedom
of the mutually interacting systems (here : of the system $A$, and
the "whole", $S_i+E_i$). It particularly means that the 
nonseparable interaction $\hat H_{S_iE_i}$ would thus remain 
intact.

On the other side, {\it and this proves RC}, neither the nonseparable
interaction $\hat H_{A(S_i+E_i)}$ could change the nonseparability
of $\hat H_{S_i+E_i}$. The proof of this assertion is as 
follows.

According to the subsection 3.1, the nonseparable interaction
$\hat H_{A(S_i+E_i)}$, if possible at all, and if meaningfull, 
would imply redefining of the complete system $(S_i+E_i) +A$,
but {\it not only (as desired)} the redefining of the system $S_i+E_i$.
This would (cf. Section 2) make the system $A$ indivisable
from the system $S_i+E_i$. But this produces a contradiction.
Actually, indistinguishability of the system $A$ contradicts
the separability of the interaction $\hat H_{AE_A}$. The only way
to overcome this contradiction is to be concerned with another the
"whole", $(S_i+E_i)+A+E_A$ (instead of the system $(S_i+E_i)+A$).
But this is nothing else but extending the original task (which
refers to $S_i+E_i$), onto the new "whole" ($(S_i+E_i)+A+E_A$),
which {\it proves imposibility of the change of the nonseparable 
(separable) local interaction to the separable (nonseparable) 
interaction
via an action from outside}.

\bigskip

{\bf 3.3 A new physical role of the separability}

\medskip

Besides giving the proof of RC, the above subsections provide
us with some interesting ideas. The separability condition 
concerning the two local interactions, $\hat H^{(i)}_{int}$ and
$\hat H^{(j)}_{int}$ cannot be considered mutually independent
(as one would plausibly expect).
should be mutually USAGLASENE. That is,
given $\hat H^{(i)}_{int}$, an interaction $\hat H^{(j)}_{int}$
cannot be of a completely arbitrary type.

This assertion follows from the previous subsection. First,  the 
separability of $\hat H^{(i)}_{int}$ simultaneously defines the "system"
$S_i$ and its environment $E_i$. According to Eq. (4), there are
the interactions between the "systems", e.g., $S_i$ and $S_j$
(likewise the interactions of their "environents"). The interaction
between the two systems is defined by their degrees of freedom.
But the degrees of freedom of $S_j$ are determined by the condition
of separability of $\hat H^{(j)}_{int}$ - which gives {\it the connection 
of the two local interactions},  $\hat H^{(i)}_{int}$
and  $\hat H^{(j)}_{int}$. It is interesting to note that these
"connections" are transitive, thus leading to a new physical
picture : the overall condition of the separability in a 
system consisting in many composite systems, assumes that
the separability conditions of the
local interactions are mutually tied, 
thus giving a consistent physicsl picture
in the EISR theory - which is only plausibly and poorely stated by the
"requirement for classicality" [2].

In general, everything told in this Section refers to a particular instant
of time, $t$. The discussion concerning this issue here will be
left out.

\bigskip

{\bf 4. Discusion}

\bigskip

Without any assumption coming from the outside of the separability 
considerations, we have obtaioned a {\it new physical role
of the separability in the EISR theory}: (A) The condition of
the separability of $\hat H_{int}$ may serve as a criterion
for {\it simultaneous}
defining of the "system" and of 
its "environment" (i.e., "system" is
not defined unless it is simultaneously defined with its "environment"),
and (B) In a set of the composite macroscopic systems, $S$, the
separability conditions concerning the local interactions are
mutually tied (i.e., the separability on one place
determines the separability condition on another, "distant" place
in the system $S$).
Thus the points (A) and (B) represent an elaborated form of the only
plausibly (and poorely) formulated the "requirement for
classicality" [2]. 

The point (B) allows for extending the considerations to the 
complete system $S$ ($S = \cup_i (S_i + E_i) = \cup_i S_i
+ \cup_i E_i$) as an isolated system. Then one may refer to the
system $S$ as to the "macroscopic piece of the Universe", for
which the point (B) establishes consistency and "rigidness"
of the definitions of its parts ($S_i$s and $E_i$s), which can be
considered as a counterpart of the "conditions of consistency" in
the cosmological considerations [7]. 

Therefore, for the purpose of defining "system", within the EISR
theory appears the next task: For a given set of the degrees of freedom, 
the separability of a given $\hat H_{int}$ should be tested.
If it would prove nonseparable, one should look for the cannonical 
transformations of the sort "(b)", so as to provide (if possible at all) 
a separable form of $\hat H_{int}$, thus obtaining a definition of the 
new composite system. This is really the {\it inverse task of the
EISR theory}, in which (likewise in the measurement theory), one 
constructs $\hat H_{int}$ for an {\it a priori} given set of the
degrees of freedom.

However, this procedure needs not to lead to unique result. 
Actually, in general one may obtain the different results with
respect to the next criteria : (i) if there appear (at least) the two
different systems ($S_1, S_2$) and their environments ($E_1, E_2$),
both refering to the separable forms of $\hat H_{int}$, and (ii) even 
for unique result (unique composite system $S+E$), one may pose the question 
of the choice of the cannonical variables describing the "system" 
(and its "environment") - which is even more difficult proiblem of
"what is 'object'?" [8].
Therefore, the above mentioned task is extended by the tasks corresponding
to the points (i) and (ii). Yet, the elaboration of these tasks
depends on the details of the model of $\hat H_{int}$, and here will be
left out.

So far, we have been concerned only with the interaction Hamiltonian, without
taking into accountr the other terms of the complete Hamiltonian (Eq. (4)).
Certainly, so as to make the results of the above tasks complete,
one must apply the same method (and reasoning) to the complete
Hamiltonian. {\it Only in this way} one may obtain the fully sensible 
definition of "system" (and of its "environment").

Finally, one may doubt about existence of the cannonical transformations
Eq. (8a-d), which should provide the transition from the nonseparable
(separable), to the separable (nonseparable) form of the Hamiltonian. 
Again, this is rather a matter of the details in the model of the 
Hamiltonian, but for an example see Appendix II.

\bigskip

{\bf 5. Conclusion}

\bigskip

We have investigated the physical meaning of the separability
of the interaction Hamiltonian. Actually, we have confronted the
separability as a necessary condition for decoherence [5, 6], 
with the "requirement
for classicality" [2], thus obtaining some interesting results.

When expressed in terms of the concept of the separability, the
"requirement for classicality" can be proved ; i.e., this plausible
statement [2] appears as a corollary of the separability considerations
in the context of the decoherence (EISR) theory. This way comes to
scope a {\it physical role of the separability}. Particularly,
the condition of the separability may serve as a criterion for
defining "system", but so that the "system" is not defined
unless it is simultaneously defined by its "environment". Second, in 
a set of mutually interacting open quantum systems, one meets 
mutual connections of the separability conditions concerning the
local interactions. The later gives a physically richer formulation
of the "requirement for classicality", which is otherwise only
a plausible statement.

\bigskip

{\bf Appendix I}

\medskip

In Appendix II of Ref. [6] it was emphasized that the necessary conditions 
might break in some exceptional cases - the existence of which has not 
been proved
but just not disproved (cf. Ref. [9]). 
Particularly, for some special models
of the interaction Hamiltonian, {\it and} for some special the initial 
states of the environment, one  eventually 
might obtain the occurrence of decoherence
even if the necessary conditions are not fulfiled. However, one can
forget about these exceptions, for many reasons. Probably the most striking
one is the next one: a special choice of the initial state of the  
environment {\it requires the preparation} of the initial state. 
But then remains the question:
who would provide this preparation? And the answer can be stated by
making reference to Omn\' es [8]: 
that one can not consider the environment's
environment ("apparatus" acting on the environment) physically sound idea,
thus removing the problem of the preparation (and also the question of the
special choice) of the initial state of the environment.

\bigskip

{\bf Appendix II}

\bigskip

We are interesting in the next two problems: First, whether the cannonical
transformations (8a-d) can provide the transformation of $\hat H_{int}$
of a nonseparable, to a separable form; Second (cf. the task (i) in Section
4), whether there exist the two different, {\it both separable}, forms
of $\hat H_{int}$, which correspond to the different composite
systems, $S_1+E_1$, and $S_2+E_2$?

In answering, we shall first refer to the second question by analyzing
the {\it interaction Hamiltonian}. Then we shall refer to the first 
question, but by analyzing the {\it complete Hamiltonian}.

Let us consider the {\it hydrogen atom} Hamiltonian :
$$\hat H = {\hat {\vec p_p^2} \over 2m_p} + 
{\hat {\vec p_e^2} \over 2m_e} + V_{Coul.}, \eqno (II.1)$$

\noindent
where the subscript "p" denotes the proton and "e" denotes the 
electron, while $V_{Coul.}$ represents the Coulomb interaction. Therefore,
the interaction Hamiltonian for this "composite system", 
"electron + proton" is:
$$\hat H_{int} \equiv V_{Coul.}. \eqno (II.2)$$

As it can be easily proved (cf. point (ii) of Section 2), 
this interaction is of the separable kind
with respect to the  noncorrelated
basis $\vert \vec r_p\rangle \otimes \vert \vec r_e\rangle$. 
But, as it is probably obvious, $\hat H_{int}$ is of the separable
kind also with respect to the "center of mass", and the "relative
particle" degrees of freedom, $\vec R_{CM}$, $\vec r_{rel}$, respectively.
That is, $\hat H_{int}$ is of the separable kind also with respect to
the noncorrelated basis $\vert \vec R_{CM}\rangle \otimes 
\vert \vec r_{rel}\rangle$. Note: here one meets the two different
"composite systems", "proton + electron ($S_1+S_2$)", 
and "center of mass + the 
relative particle ($S_2+E_2$)", {\it 
both referring to the separable forms} of $\hat H_{int}$;
this is the answer to the above the second question.

However, as it was strongly emphasized in Section 5, in defining 
"system" one must take into considerations 
the {\it complete Hamiltonian},
which in this case, leads to unique definition of "system".

Actually, there are the two different forms of the complete
Hamiltonian :
$$\hat H = {\hat {\vec p_p^2} \over 2m_p} \otimes \hat I_e+ 
\hat I_p \otimes {\hat {\vec p_e^2} \over 2m_e} -
{Z e^2 \over 4\pi \epsilon_{\circ} \vert \vec r_p -
\vec r_e\vert}, \eqno (II.3)$$

\noindent
and
$$\hat H = {\hat {\vec P_{CM}^2} \over 2M} \otimes 
\hat I_{rel}
+ \hat I_{CM} \otimes \left( {\hat {\vec p_{rel}^2} 
\over 2\mu}
- {Z e^2\over 4\pi \epsilon_{\circ} \hat r_{rel}}\right),
\eqno (II.4)$$

As it easily follows from the "operational definition"
of the separability (cf. point (iv) of Section 2), the
expression (II.3) is of the nonseparable kind, while
the expression (II.4) is of the separable kind, thus
uniquely defining the composite system: it is the system
"center of mass + the relative particle", {\it defined by the separable 
form} of the complete Hamiltonian, Eq. (II.4).

Note: the cannonical transformations, $(\vec r_p, \vec p_p;
\vec r_e, \vec p_e)$ $\rightarrow (\vec R_{CM},$ $\vec P_{CM};$
$\vec r_{rel},$ $\vec p_{rel})$, provide the 
{\it transformation} of
the complete Hamiltonian from the nonseparable (Eq. (II.3)), to
the separable form (Eq. (II.4)). (In the position representation
this reads : $\hat H$ is of the nonseparable form with respect to 
a noncorrelated basis $\{\Psi_m(\vec r_p) \otimes \chi_n(\vec r_e) \}$,
but is of the separable form in a noncorrelated
basis $\{\Phi_p(\vec R_{CM})
\otimes \phi_q(\vec r_{rel})\}$.) This gives the answer to the
above the first question.

Although this model does not refer to the decoherence theory,
we feel it paradigmatic for the considerations of the 
many-particle quantum systems for which the results
strongly depend upon the details in the model of the Hamiltonian.

\vfill\eject

{\bf References :}

\item{[1]}
W. H. Zurek, Phys. Rev. {\bf D26} (1982) 1862 

\item{[2]}
W. H. Zurek, Prog. Theor. Phys. {\bf 89} (1993) 281

\item{[3]}
S. Machida and M. Namiki, in Proc. 2nd Int. Symp. Foundations
of Quantum Mechanics, Tokyo, 1986, pp. 355-359; and references therein

\item{[4]}
A. O. Caldeira and A. J. Leggett, Ann. Phys. (N.Y.) {\bf 149} 
(1983) 374

\item{[5]}
M. Dugi\' c, Physica Scripta {\bf 53} (1996) 9

\item{[6]}
M. Dugi\' c, Physica Scripta {\bf 56} (1997) 560

\item{[8]}
R. Omn\' es, "The Interpretation of Quantum Mechanics",
Princeton University Press, Princeton, 1994

\item{[9]}
M. Dugi\' c, J. Res. Phys. {\bf 27} (1998) 141.

\end